\documentclass[amsmath, aps]{revtex4}
\usepackage{times}
\usepackage{epsfig}
\usepackage{amssymb}

\usepackage{amsmath}

\begin{document}
\title{Spin-Statistic Selection Rules for Multi-Equal-Photon Transitions in Atoms: Extension of the Landau-Yang Theorem to Multiphoton Systems}
\author{T. Zalialiutdinov$^{1}$, D. Solovyev$^{1}$, L. Labzowsky$^{1,2}$ and G. Plunien$^{3}$}

\affiliation{$^1$ Department of Physics, St.Petersburg State University,
Ulianovskaya 1, Petrodvorets, St.Petersburg 198504, Russia
\\
$^2$  Petersburg Nuclear Physics Institute, 188300, Gatchina, St.
Petersburg, Russia
\\
$^3$ Institute f\"{u}r Theoretische Physik, Technische Universit\"{a}t Dresden, Mommsenstrasse 13, D-10162, Dresden, Germany}

\begin{abstract}
We establish the existence of spin-statistic selection rules (SSSR) for multi-equal-photon transitions in atomic systems. These selection rules are similar to those for systems of many equivalent electrons in atomic theory. The latter ones are the direct consequence of Pauli exclusion principle. In this sense the SSSR play the role of the exclusion principle for photons: they forbid some particular states for the photon systems. We established several SSSR for few-photon systems. 1) First rule (SSSR-1): two-equivalent photons involved in any atomic transition can have only even values of the total angular momentum $ J $. This selection rule is an extension of the Landau-Yang theorem to the photons involved in atomic transitions. 2) second rule (SSSR-2): three equivalent dipole photons involved in any atomic transition can have only odd values of the total angular momentum $ J=1,3 $. 3) third rule (SSSR-3): four equivalent dipole photons involved in any atomic transition can have only even values of the total angular momentum $ J=0,2,4 $. We also suggest a method for a possible experimental test of these SSSR by means of laser experiments.
\end{abstract}

\maketitle

\section{Introduction}
Landau-Yang theorem \cite{landau}, \cite{yang}, together with the Bose-Einstein condensation can be viewed as the most spectacular confirmations of Bose-Einstein statistics for integer-value-spin particles. Landau-Yang theorem forbids two photons to participate in any process that would require them to be in a state with total angular momentum one. In the high-energy physics an evident example is the prohibition the two-photon decay for the neutral spin-one $Z^{0}$-boson. The same concerns also the annihilation decay of orthopositronium (also spin-1 state). However both this decays are also forbidden by charge-parity conservation law. Positronium presents a real neutral system (it coincides with itself after charge conjugation), therefore it possesses a definite charge parity \cite{berlif}, connected with the total spin value $S$: parapositronium ($S=0$) is charge-positive and orthopositronium ($ S=1 $) is charge negative. Since the charge parity of a system of $ N_{\gamma} $ photons equals $ (-1)^{N_{\gamma}} $ \cite{akhiezer}, parapositronium can not decay into an odd number of (not necessarily equivalent) photons and orthopositronium can not decay into an even number  of photons. The $ Z^{0} $-boson as a charge-parity-negative particle can not decay into an even number of photons.

A similar situation exists in atomic physics. Already the early calculations of the two-photon decay of the singlet $ 2^1S_0\equiv(1s2s)^1S_0 $ and triplet $ 2^3S_1\equiv(1s2s)^3S_1 $ excited states of the He-like ions to the ground $ 1^1S_0\equiv(1s)^{2\;1}S_0 $ state revealed the crucial difference in the photon frequency distributions in both cases \cite{4a}. The decay probability for the triplet case tends to zero when the frequencies of the emitted photons are equal (see Fig. 1). Later these conclusions were confirmed within the fully relativistic calculations (see, for example \cite{derevianko}).

Unlike the positronium two-electron ions do not possess a definite charge parity. The neutral He atom also is not a real neutral particle and also does not have a definite charge parity. Therefore, only the spin-statistic properties can be responsible for this special selection rule. Its connection with the Landau-Yang theorem and hence with Bose-Einstein statistics was first emphasized in \cite{budker} where an experimental limit for the violation of the Spin-Statistic Theorem (SST) was obtained. Recently this limit was improved in \cite{english}. A confirmation of the Spin-Statistic-Selection Rules (SSSR) for two-equal-photon transitions in atomic physics was obtained in \cite{dunford} where it was demonstrated  that the frequency distributions for the transition rates $ 2^3S_1\rightarrow 1^1S_0+2\gamma $ with $ 2E1 $, $ 2M1 $, $ 2E2 $ photons correspond to the type Fig. 1 (right panel), while the distributions for the same transition with $ E1M2 $, $ E2M1 $ (nonequivalent) photons belong to the type Fig. 1 (left panel). Finally in \cite{kozlov_eng} the SSSR-suppressed two-photon transition induced either by the hyperfine interaction or by an external magnetic field were investigated. The hyperfine interaction changes the value of the total angular momentum $ J_e $ of electron state in an atom and in an external magnetic field the angular momentum, in principle, is not conserved. Therefore in both cases the prohibition of the atomic transition $ J_e=1\rightarrow J_e=0 $ with two equivalent photons can be relaxed. 

We formulate the SSSR for the multi-equal-photon atomic transitions which present an extension of the Landau-Yang theorem as follows:

1) SSSR-1: Two equivalent photons involved in any atomic transition can have only even values of the total angular momentum $J$,

2) SSSR-2: Three equivalent dipole photons involved in any atomic transition can have only odd values of the total angular momentum $J=1, \,3$,

3) SSSR-3: Four equivalent dipole photons involved in any atomic transition can have only even values of the total momentum values $J=0,\,2,\, 4$.

Note that SSSR-2, SSSR-3 do not hold, in general, for the photon multipolarity $j>1$ (see sections IV, V).

The Landau-Yang theorem in case of atomic transitions corresponds to SSSR-1 for number of photons $N_{\gamma}=2$ and multipolarity $J=1$ involved in the transitions $J_e=1\rightarrow J_e=0$ or $J_e=0 \rightarrow J_e=1$ via the emission or absorption of two photons. In principle, the original Landau-Yang theorem does not require that the two photons are equivalent; it is assumed only that they are not propagating in the same direction, so that one can choose the frame of reference where the center-of-inertia for two photons is at rest. Moreover, originally Landau-Yang theorem was applied to the processes when the initial particle disappears after decay and is converted to a system of two photons. From the energy-momentum conservation law it follows that these two photons should be collinear (opposite direction) and should have equal frequencies. The total angular momentum and space parity for these photons were defined in the center-of-inertia rest frame  for the two photons. This frame coincides with the rest frame for decaying particle which disappears after decay ($ Z^{0} $-boson, annihilation decay of orthopositronium). For introducing of the SSSR for two-photon atomic decays we employ a different picture (see section II). We define the equivalent photons as the photons having the same frequency, angular momentum and parity in the rest frame of a decaying atom. In atomic processes the decaying particle (atom) does not disappear in the decay process, therefore the rest frame of an atom does not coincide with the center-of-inertia rest frame for two emitted photons. The difference in the definitions of the total angular momenta and space parity in both reference frames is important. The value of the photon orbital angular momentum and therefore the value of the photon total angular momentum depends on the choice of the reference frame. 

For all these reasons the SSSR-1 which will be proven explicitly in section II for the two-photon decay does not fully coincide with the results of the standard Landau-Yang theorem for the two-photon decay of disappearing particle \cite{landau}-\cite{akhiezer}. In the standard formulation a two-photon system after decay of an initial (disappearing) particle can have any total angular momentum $J$, except $J=1$.

This result was obtained in the center-of-inertia rest frame for two photons and is not immediately applicable to an arbitrary two-photon decays in He-like ions. According to the SSSR-1 no odd values for the total angular momentum are allowed. Thus our SSSR-1 in case of two equivalent photons is more restrictive than one could expect from the standard formulation of the Landau-Yang theorem.

The standard formulation of Landau-Yang theorem can not be applied also to laser photons: these photons are collinear but have the same direction and the center-of-inertia rest frame does not exist in this case. With our approach we can consider the absorption of the laser photons in the rest frame of an absorbing atom or ion. The SSSR will work for the absorption transition as well. Though the incident laser photons do not have certain total angular one-photon momentum in the rest frame of an absorbing atom, the fixing of the photon frequency usually defines the initial and final atomic states in the process of absorption. The total electron angular momenta $ J_e $ of the initial and final atomic states then define the total angular momentum for a photon which can be absorbed in this particular transition. In case of multiphoton absorption the total angular momenta of the absorbed photons are defined by the vector coupling scheme.

Apart from the general proof of SSSR-1 in case of two-photon decay we support this proof by the direct evaluation of the frequency distribution of transition rate in section II for the transition $J_e=3\rightarrow J_e=0+2\gamma(E2)$, i.e. for the case when the 2-photon system has a total angular momentum $J=3$. The transition rate tends to zero when the frequency of both photons are equal.

In section III we search for the analogy between the values of a total angular momentum for the system of equivalent photons (allowed by the SST) and the values of a total angular momentum for the system of equivalent atomic electrons also allowed by SST, only with the Fermi-Dirac statistics. Section IV contains a detailed derivation of the 3-photon transition rate in Highly-Charged Ions (HCI) and the proof of SSSR-2 for the particular case of 3-photon transition $J_e=2 \rightarrow J_e=0 +3\gamma(E1)$. The results for the other 3-dipole-photon transitions are also included in this section. Section V contains the analytic proof of the SSSR-2 with equations for the Coefficients of Fractional Parentage traditionally used in the theory of atomic spectra for constructing the wave functions for equivalent electrons. In section VI the analytic proof is given dor SSSR-3 and the particular examples are provided for the 4-photon transitions which support this proof. Section VII contains discussion and outlook. A possible experimental test of the results presented in this paper is briefly discussed.

\section{Proof of the SSSR-1 for two-photon transitions}

We describe a photon emitted or absorbed by an atom by wave functions $\vec{A}^{(s)*}_{\omega jm}(\vec{k})$ or $\vec{A}^{(s)}_{\omega jm}(\vec{k})$ in momentum space, respectively \cite{berlif}. Here $\omega$ is in the frequency, $jm$ are the angular momentum and its projection and index $s$ denotes the type of the photon - electric ($s=E$) or magnetic $(s=M)$. The type of the photon together with $j$ value determines the parity $P$ of the photon: $P=(-1)^{j+1}$ for $s=E$ or $P=(-1)^{j}$ for $s=M$; the argument $\vec{k}$ denotes the photon momentum. We denote also the vector component of the photon wave function as $\left(\vec{A}^{(s)}_{\omega jm}\right)_i$ where the index $i=1,2,3$. For the real transverse $E,M$ photons the index $i$ takes only two values $i=1,2$, while $i=3$ corresponds to the longitudinal component which is absent for $E,M$ photons. Each component of these wave functions is the eigenstate of the total one-photon angular momentum operator
\begin{eqnarray}
\label{1}
\widehat{\vec{j}}^{2}\left(\vec{A}^{(s)}_{\omega jm}\right)_i=j(j+1)\left(\vec{A}^{(s)}_{\omega jm}\right)_i,
\end{eqnarray}
\begin{eqnarray}
\label{2}
 \widehat{\vec{j}}_z\left(\vec{A}^{(s)}_{\omega jm}\right)_i=m\left(\vec{A}^{(s)}_{\omega jm}\right)_i.
\end{eqnarray}

The two-photon wave function for two photons with the same frequency can be constructed as a symmetrized tensor product
\begin{eqnarray}
\label{3}
\left(\Phi^{s_1s_2}_{\omega JM}(\vec{k}_1\vec{k}_2)\right)_{i_1,i_2} = N\sum_{m_1m_2}C^{JM}_{j_1m_1j_2m_2}\left[\left(\vec{A}^{(s_1)}_{\omega j_1m_1}(\vec{k}_1)\right)_{i_1}\left(\vec{A}^{(s_2)}_{\omega j_2m_2}(\vec{k}_2)\right)_{i_2} + 
\left(\vec{A}^{(s_1)}_{\omega j_1m_1}(\vec{k}_2)\right)_{i_2}\left(\vec{A}^{(s_2)}_{\omega j_2m_2}(\vec{k}_1)\right)_{i_1}\right].
\end{eqnarray}
Here indices 1, 2 correspond to the two photons, $JM$ are the total angular momentum for the two-photon system and its projection; one of the standard notations for the Clebsh-Gordan coefficient $C^{JM}_{j_1m_1j_2m_2}$ is used. An explicit expression for the normalization factor $N$ is not necessary for our purposes. The components of the tensor wave function (\ref{3}) are the eigenstates of the total two-photon angular momentum operator
\begin{eqnarray}
\label{4}
\left(\widehat{\vec{j}}_1+\widehat{\vec{j}}_2\right)^2\left(\Phi^{(s_1s_2)}_{\omega JM}\right)_{i_1i_2}=J(J+1)\left(\Phi^{(s_1s_2)}_{\omega JM}\right)_{i_1i_2}
\end{eqnarray}
and its projection
\begin{eqnarray}
\label{5}
\left(\widehat{j}_{1z}+\widehat{j}_{2z}\right)\left(\Phi^{(s_1s_2)}_{\omega JM}\right)_{i_1i_2}=M\left(\Phi^{(s_1s_2)}_{\omega JM}\right)_{i_1i_2}.
\end{eqnarray}
Bose-Einstein symmetry is already implemented in Eq. (\ref{3}) via symmetrization. For the equivalent photons $j_1=j_2=j$ and $s_1=s_2=s$. Then, changing the notations for the summation indices $m_1\leftrightarrows m_2$ in the second term in square brackets in Eq. (\ref{3}) we can rewrite this equation like
\begin{eqnarray}
\label{6}
\left(\Phi^{s\;s}_{\omega JM}(\vec{k}_1\vec{k}_2)\right)_{i_1,i_2}=N\sum_{m_1m_2}\left(\vec{A}^{(s)}_{\omega jm_1}(\vec{k}_1)\right)_{i_1}\left(\vec{A}^{(s)}_{\omega jm_2}(\vec{k}_2)\right)_{i_2}\left[C^{JM}_{jm_1jm_2}+C^{JM}_{jm_2jm_1}\right].
\end{eqnarray}
Employing the symmetry properties for the Clebsh-Gordan coefficients (with integer $j_1,j_2$):
\begin{eqnarray}
\label{7}
C^{JM}_{j_2m_2j_1m_1}=(-1)^{j_1+j_2+J}C^{JM}_{j_1m_1j_2m_2}
\end{eqnarray}
we obtain finally an expression for the wave function of two equivalent photons
\begin{eqnarray}
\label{8}
\left(\Phi^{s\;s}_{\omega JM}(\vec{k}_1\vec{k}_2)\right)_{i_1,i_2}=N\left[1+(-1)^{2j+J}\right]\sum_{m_1m_2}\left(\vec{A}^{(s)}_{\omega jm_1}(\vec{k}_1)\right)_{i_1}\left(\vec{A}^{(s)}_{\omega jm_2}(\vec{k}_2)\right)_{i_2}.
\end{eqnarray}
This wave function vanishes for the odd values of $ J $. Thus, SSSR-1 prohibits all the odd values of $ J $ for two equivalent photons and, consequently the corresponding transitions. According to Eq. (\ref{8}) the two-photon  wave  function for two equivalent photons always is of even parity. We should note that description of the properties of a two-photon wave function, close to presented here can be found also in \cite{9a}, as well as similar to our treatment of the Landau-Yang theorem in atomic processes.

The SSSR-1 for $2^3S_1\rightarrow 1^1S_0+2\gamma(E1)$ two-equal-frequency-photon transition in He-like ions follows directly from Eq. (\ref{8}). To illustrate  SSSR-1 more directly and to support our analytical proof we performed also the evaluation of the transition rate $3^3D_3\equiv (3d1s)^3D_3\rightarrow 1^1S_0+2\gamma(E2)$ for He-like U ($Z=92$). This evaluation is fully similar to evaluation of $2^3S_1\rightarrow 1^1S_0+2\gamma(E1)$ transition rate (see capture to Fig. 1). Photon frequency distribution for the two-photon $3^3D_3\rightarrow 1^1S_0+2\gamma E(2)$ transition is given Fig. 2. The results in Fig. 2 demonstrate that SSSR-1 prohibits for two equivalent photons emitted in atomic transition to have a total angular momentum $J=3$. As it was explained in the Introduction, it does not contradict to the "standard" formulation of the Landau-Yang-theorem which allows this value for the two-photon decay of the particle provided that this particle disappears in the process of the decay and the rest system for the center-of-inertia for two photons is employed.

\section{SSSR for equivalent electrons in atoms: comparison with the SSSR for multi-equal-photon transitions}

There exists an analogy between the total electron momentum $J_e$ values allowed by the corresponding SSSR within $jj$-coupling scheme (in this case the spin-statistics is of the Fermi-Dirac type) and the total photon momentum values $J$ allowed by Bose-Einstein SSSR. The one-electron Dirac wave functions in coordinate space  we denote like $\psi_{nj_el_em_e}(\vec{r})$, where $n$ is the principal quantum number, $j_e,m_e$ are the electron total angular momentum and its projection, $l_e$ is the orbital angular momentum value which defines the parity of the state $P=(-1)^{l_e}$, $\vec{r}$ is the space coordinate. In what follows we will omit the spinor indices. The two-electron atomic wave function (a 16-component spinor) having definite total angular momentum $J_e$ and its projection $M_e$ can be constructed as
\begin{eqnarray}
\label{9}
\psi^{(l_{e_1}l_{e_2})}_{n_1n_2J_eM_e}(\vec{r}_1,\vec{r}_2)=N_e
\sum_{m_{e_1}m_{e_2}}C^{J_eM_e}_{j_{e_1}m_{e_1}j_{e_2}m_{e_2}}\left[\psi_{n_1j_{e_1}l_{e_1}m_{e_1}}(\vec{r}_1)\psi_{n_2j_{e_2}l_{e_2}m_{e_2}}(\vec{r}_2)-\psi_{n_1j_{e_1}l_{e_1}m_{e_1}}(\vec{r}_2)\psi_{n_2j_{e_2}l_{e_2}m_{e_2}}(\vec{r}_1)\right],
\end{eqnarray}
where $N_e$ is the normalization factor. The Fermi-Dirac statistics is implemented in Eq. (\ref{5}) via the antisymmetrization. The parity of the two-electron wave function is defined as $P=(-1)^{l_{e_1}+l_{e_2}}$. For the equivalent electrons (i.e. the electrons from the one nonclosed shell) $n_1=n_2=n$, $j_{e_1}=j_{e_2}=j_e$, $l_{e_1}=l_{e_2}=l_e$. Then, using the same procedure as in section II, we obtain 
\begin{eqnarray}
\label{10}
\psi^{(l_el_e)}_{nnJ_eM_e}(\vec{r}_1,\vec{r}_2)=N_e\left[1-(-1)^{2j_e+J_e}\right]\sum_{m_{e_1}m_{e_2}}C^{J_eM_e}_{j_ej_em_{e_1}m_{e_2}}\psi_{n
j_{e}l_{e}m_{e}}(\vec{r}_1)\psi_{nj_{e}l_{e}m_{e}}(\vec{r}_2).
\end{eqnarray}
The parity of this function is always even as for photon wave function Eq. (\ref{8}). From Eq. (\ref{10}) it follows that the SSSR allows for two equivalent electrons also only even values of $J_e$, exactly as for the equivalent photons. This happens because in the second term in square brackets in Eq. (\ref{10}) we have an additional factor $ (-1) $ unlike in Eq. (\ref{8}), while the values of $j_e$, unlike the values of $j$, are half-integer.

In case of the number of equivalent electrons $N_e>2$ there is no simple way to define the allowed values of the total angular momentum $J_e$. To determine these values one has to write down all sets of projections $m_{e_1},\dots m_{e_{N_e}}$ which do not violate the Pauli principle. Then for each set the total projection $M_e$ should be defined and all $M_e$ values should be distributed between possible values of $J_e$. This is a lengthy procedure (for example, for $j_e=7/2$ and $N_e=4$ the number of such sets equals to 70) which can be only partly simplified with the use of group theory. The results can be found in books on atomic spectroscopy (for example \cite{sobelman}) and are presented in Table 1. In general, all numbers in this Table follow from the Pauli principle. However, if we apply SSSR-1, SSSR-3, etc. for defining the allowed values of $J_e$ we will see that it will work up to $j_e=7/2$, $N_e=4$. It is violated by the presence of $J_e=5$ for $j_e=7/2$, $N_e=4$. So the SSSR-3 for the equivalent electrons is limited by the value $j_e=5/2$. This should be compared with the limitation of the SSSR-2, SSSR-3 etc. by the photon multipolarity $j=1$. 

Finally, we should note that the similarity between the systems of equivalent photons  and equivalents electrons could be made more close with the use of the matrix form of Maxwell equations \cite{mohr}.

\section{SSSR-2 for three-dipole photon transitions}

For photon numbers $N_{\gamma}>2$ the SSSR can be formulated only for dipole photons. For 3-dipole-photon transitions we can give few examples which prove the validity of SSSR-2 for $N_{\gamma}=3$. In this section we present a detailed derivation for this proof. First, we present a computationally convenient fully relativistic form of a general expression for the 3-photon decay rate in H-like ion for arbitrary combination of electric and magnetic multipoles and in an arbitrary gauge for the electromagnetic potentials.

The $S$-matrix element for the process $i\rightarrow f+3\gamma$ ($i$ and $f$ denote the initial and final states of H-like ion respectively) reads \cite{berlif}, \cite{akhiezer}, \cite{ALPS}
\begin{eqnarray}
\label{11}
S_{fi}^{(3)}=(-ie)^3\int d^4x_3d^4x_2d^4x_1\overline{\psi}_{f}(x_3)\gamma_{\mu_3}A_{\mu_3}^{*\left(\vec{k}_{3}\vec{e}_{3}\right)}(x_3)S(x_3,x_2)\gamma_{\mu_2}A_{\mu_2}^{*\left(\vec{k}_{2}\vec{e}_{2}\right)}(x_2) S(x_2,x_1)\gamma_{\mu_1}A_{\mu_1}^{*\left(\vec{k}_{1}\vec{e}_{1}\right)}(x_1)\psi_{i}(x_1),
\end{eqnarray}
\begin{eqnarray}
\label{12}
\psi_n(x)=\psi_n(\vec{r})e^{-iE_nt},
\end{eqnarray}
$\psi_n(\vec{r})$ is the solution of the Dirac equation for the atomic electron, $E_n$ is the Dirac energy, $\overline{\psi}_n=\psi^+_n\gamma_0$ is the Dirac conjugated wave function, $\gamma_{\mu}\equiv(\gamma_0,\vec{\gamma})$ are the Dirac matrices and $x\equiv(\vec{r},i\,t)$ are the space-time coordinates. In this paper the Euclidean metric with an imaginary fourth component is adopted. The photon wave function (electromagnetic field potential) is described by
\begin{eqnarray}
\label{13}
A_{\mu}^{(\vec{k},\vec{e})}(x)=\sqrt{\frac{2\pi}{\omega}}e_{\mu}e^{ik_{\mu}x_{\mu}}=A_{\mu}^{(\vec{k},\vec{e})}(\vec{r})e^{-i\omega t},
\end{eqnarray}
where $k\equiv(\vec{k},i\omega)$ is the photon momentum 4-vector, $\vec{k}$ is the photon wave vector, $\omega = |\vec{k}|$ is the photon frequency, $e_{\mu}$ are the components of the photon polarization 4-vector, $\vec{e}$ is the 3-dimensional polarization vector for real photons, $A_{\mu}^{(\vec{k},\vec{e})}$ corresponds to the absorbed photon, $A_{\mu}^{*(\vec{k},\vec{e})}$ corresponds to the emitted photon, respectively. 

For the real transverse photons
\begin{eqnarray}
\label{14}
\vec{A}(x)=\sqrt{\frac{2\pi}{\omega}}\vec{e}e^{i(\vec{k}\vec{r}-\omega t)}\equiv\sqrt{\frac{2\pi}{\omega}}\vec{A}_{\vec{e},\vec{k}} e^{-i\omega t}.
\end{eqnarray}

The electron propagator for bound electrons we present in the form of the eigenmode decomposition with respect to one-electron eigenstates \cite{berlif}, \cite{akhiezer}
\begin{eqnarray}
\label{15}
S(x_1,x_2)=\frac{1}{2\pi i}\int\limits^{\infty}_{-\infty}d\omega e^{i\omega(t_1-t_2)}\sum\limits_{n}\frac{\psi_n(\vec{r}_1)\overline{\psi}_n(\vec{r}_2)}{E_n(1-i0)+\omega},
\end{eqnarray}
where summation is runs over entire Dirac spectrum for atomic electron. Insertion of the expressions (\ref{12})-(\ref{15}) into Eq. (\ref{11}) and performing the integrations over time and frequency variables yields
\begin{eqnarray}
\label{16}
S_{fi}^{(3)}=-2\pi i e^3\delta(E_i-E_f-\omega_3-\omega_2-\omega_1)\sum\limits_{n'n}\frac{\left(\vec{\alpha}\vec{A}^*_{\vec{e}_3,\vec{k}_3}\right)_{fn'}\left(\vec{\alpha}\vec{A}^*_{\vec{e}_2,\vec{k}_2}\right)_{n'n}\left(\vec{\alpha}\vec{A}^*_{\vec{e}_1,\vec{k}_1}\right)_{ni}}{(E_{n'} - E_f - \omega_3) (E_n - E_f - \omega_3 - \omega_2)},
\end{eqnarray}
where $\vec{\alpha}$ are the Dirac matrices, $(\dots)_{kn}$ denotes the matrix element with Dirac wave function $\psi_k$, $\psi_n$. The amplitude $U$ of the process is related to the S-matrix via
\begin{eqnarray}
\label{17}
S_{fi}=-2\pi i \delta(E_i-E_f-\omega_3-\omega_2-\omega_1)U_{fi}.
\end{eqnarray}
The probability (differential transition rate) of the process is defined as
\begin{eqnarray}
\label{18}
\frac{dW^{3\gamma}_{i\rightarrow f}}{d\omega_3d\omega_2d\omega_1}=2\pi\delta(E_i-E_f-\omega_3-\omega_2-\omega_1)\left|U^{(3)}_{fi}\right|.
\end{eqnarray}
We will be interested in the expression for the transition rate integrated over the directions $ \vec{\nu}=\vec{k}/|\vec{k}| $ and summed over the photon polarizations $ \vec{e} $ of all the emitted photons. Then, taking into account all the permutations of photons and integrating over $\omega_3$, we find
\begin{eqnarray}
\label{19}
\frac{dW_{i\rightarrow f}(\omega_1,\omega_2)}{d\omega_1d\omega_2} = \frac{\omega_3\;\omega_2\;\omega_1}{(2\pi)^5}\sum_{\vec{e}_3,\vec{e}_2,\vec{e}_1}\int d\vec{\nu}_1d\vec{\nu}_2d\vec{\nu}_3 \times
\\
\nonumber
\left| \sum\limits_{n'n}\frac{\left(\vec{\alpha}\vec{A}^*_{\vec{e}_3,\vec{k}_3}\right)_{fn'}\left(\vec{\alpha}\vec{A}^*_{\vec{e}_2,\vec{k}_2}\right)_{n'n}\left(\vec{\alpha}\vec{A}^*_{\vec{e}_1,\vec{k}_1}\right)_{ni}}{(E_{n'} - E_f - \omega_3) (E_n - E_f - \omega_3 - \omega_2)}\right.+
\sum\limits_{n'n}\frac{\left(\vec{\alpha}\vec{A}^*_{\vec{e}_3,\vec{k}_3}\right)_{fn'}\left(\vec{\alpha}\vec{A}^*_{\vec{e}_1,\vec{k}_1}\right)_{n'n}\left(\vec{\alpha}\vec{A}^*_{\vec{e}_2,\vec{k}_2}\right)_{ni}}{(E_{n'} - E_f - \omega_3) (E_n - E_f - \omega_3 - \omega_1)}+
\\
\nonumber
\sum\limits_{n'n}\frac{\left(\vec{\alpha}\vec{A}^*_{\vec{e}_1,\vec{k}_1}\right)_{fn'}\left(\vec{\alpha}\vec{A}^*_{\vec{e}_3,\vec{k}_3}\right)_{n'n}\left(\vec{\alpha}\vec{A}^*_{\vec{e}_2,\vec{k}_2}\right)_{ni}}{(E_{n'} - E_f - \omega_1) (E_n - E_f - \omega_1 - \omega_3)}+
\sum\limits_{n'n}\frac{\left(\vec{\alpha}\vec{A}^*_{\vec{e}_1,\vec{k}_1}\right)_{fn'}\left(\vec{\alpha}\vec{A}^*_{\vec{e}_2,\vec{k}_2}\right)_{n'n}\left(\vec{\alpha}\vec{A}^*_{\vec{e}_3,\vec{k}_3}\right)_{ni}}{(E_{n'} - E_f - \omega_1) (E_n - E_f - \omega_1 - \omega_2)}+
\\
\nonumber
\sum\limits_{n'n}\frac{\left(\vec{\alpha}\vec{A}^*_{\vec{e}_2,\vec{k}_2}\right)_{fn'}\left(\vec{\alpha}\vec{A}^*_{\vec{e}_3,\vec{k}_3}\right)_{n'n}\left(\vec{\alpha}\vec{A}^*_{\vec{e}_1,\vec{k}_1}\right)_{ni}}{(E_{n'} - E_f - \omega_2) (E_n - E_f - \omega_2 - \omega_3)}+
\left.\sum\limits_{n'n}\frac{\left(\vec{\alpha}\vec{A}^*_{\vec{e}_2,\vec{k}_2}\right)_{fm}\left(\vec{\alpha}\vec{A}^*_{\vec{e}_1,\vec{k}_1}\right)_{n'n}\left(\vec{\alpha}\vec{A}^*_{\vec{e}_3,\vec{k}_3}\right)_{ni}}{(E_{n'} - E_f - \omega_2) (E_n - E_f - \omega_2 - \omega_1)}\right|^2,
\end{eqnarray}
where $\omega_3=E_i-E_f-\omega_2-\omega_1$.
The total transition rate can be defined as 
\begin{eqnarray}
\label{21}
W_{i\rightarrow f}=\frac{1}{3!}\frac{1}{2j_i+1}\sum\limits_{M_i,M_f}\;\iint\limits_{\omega_3\geqslant 0}dW_{i\rightarrow f}(\omega_1,\omega_2).
\end{eqnarray}
Expanding the plane waves into spherical waves in Eq. (\ref{19}) we go over to the description of photons by the total angular momentum  $J$, its projection $M$ and parity (type of the photon). Then summation over polarizations and integration over photon emission angles in Eq. (\ref{19}) yields
\begin{eqnarray}
\label{20}
\frac{dW_{i\rightarrow f}(\omega_1,\omega_2)}{d\omega_1d\omega_2} = \frac{\omega_3\;\omega_2\;\omega_1}{(2\pi)^5}\sum_{\lambda_3\lambda_2\lambda_1}\sum_{J_3J_2J_1}\sum_{M_3M_2M_1}
\\
\nonumber
\left|\sum\limits_{n'n}\frac{\left(Q^{(\lambda_3)}_{J_3M_3\omega_3}\right)_{fn'}\left(Q^{(\lambda_2)}_{J_2M_2\omega_2}\right)_{n'n}\left(Q^{(\lambda_1)}_{J_1M_1\omega_1}\right)_{ni}}{(E_{n'} - E_f - \omega_3) (E_n - E_f - \omega_3 - \omega_2)}+
\sum\limits_{n'n}\frac{\left(Q^{(\lambda_3)}_{J_3M_3\omega_3}\right)_{fn'}\left(Q^{(\lambda_1)}_{J_1M_1\omega_1}\right)_{n'n}\left(Q^{(\lambda_2)}_{J_2M_2\omega_2}\right)_{ni}}{(E_{n'} - E_f - \omega_3) (E_n - E_f - \omega_3 - \omega_1)}+
\right.
\\
\nonumber
\sum\limits_{n'n}\frac{\left(Q^{(\lambda_1)}_{J_1M_1\omega_1}\right)_{fn'}\left(Q^{(\lambda_3)}_{J_3M_3\omega_3}\right)_{n'n}\left(Q^{(\lambda_2)}_{J_2M_2\omega_2}\right)_{ni}}{(E_{n'} - E_f - \omega_1) (E_n - E_f - \omega_1 - \omega_3)}+
\sum\limits_{n'n}\frac{\left(Q^{(\lambda_1)}_{J_1M_1\omega_1}\right)_{fn'}\left(Q^{(\lambda_2)}_{J_2M_2\omega_2}\right)_{n'n}\left(Q^{(\lambda_3)}_{J_3M_3\omega_3}\right)_{ni}}{(E_{n'} - E_f - \omega_1) (E_n - E_f - \omega_1 - \omega_2)}+
\\
\nonumber
\sum\limits_{n'n}\frac{\left(Q^{(\lambda_2)}_{J_2M_2\omega_2}\right)_{fn'}\left(Q^{(\lambda_3)}_{J_3M_3\omega_3}\right)_{n'n}\left(Q^{(\lambda_1)}_{J_1M_1\omega_1}\right)_{ni}}{(E_{n'} - E_f - \omega_2) (E_n - E_f - \omega_2 - \omega_3)}+
\left.\sum\limits_{n'n}\frac{\left(Q^{(\lambda_2)}_{J_2M_2\omega_2}\right)_{fn'}\left(Q^{(\lambda_1)}_{J_1M_1\omega_1}\right)_{n'n}\left(Q^{(\lambda_3)}_{J_3M_3\omega_3}\right)_{ni}}{(E_{n'} - E_f - \omega_2) (E_n - E_f - \omega_2 - \omega_1)}\right|^2,
\end{eqnarray} 
where we employ the reduction of the matrix elements to the radial integrals developed in \cite{grant}, \cite{12a}
\begin{eqnarray}
\label{22}
\left\langle n_{\alpha}j_{\alpha}l_{\alpha}m_{\alpha}|Q^{(\lambda)}_{J_{\gamma}M\omega}|n_{\beta}j_{\beta}l_{\beta}m_{\beta}\right\rangle=(-1)^{j_{\alpha}-m_{\alpha}}
\begin{pmatrix}
j_{\alpha} & J_{\gamma} & j_{\beta}\\
-m_{\alpha} & M_{\gamma} & m_{\beta}
\end{pmatrix}\times
\\
\nonumber
(-i)^{J+\lambda-1}(-1)^{j_{\alpha}-1/2}\left(\dfrac{4\pi}{2J+1}\right)^{1/2}
\left[(2j_{\alpha}+1)(2j_{\beta}+1)\right]^{1/2}
\begin{pmatrix}
j_{\alpha} & J_{\gamma} & j_{\beta}\\
1/2 & 0 & -1/2
\end{pmatrix}
\overline{M}^{(\lambda,J_{\gamma})}_{n_{\alpha}l_{\alpha}n_{\beta}l_{\beta}}.
\end{eqnarray}
Here $J_{\gamma}$, $M_{\gamma}$ are the total angular momentum of the photon and its projection, $\lambda$ characterizes the type of the photon: $\lambda=1$ corresponds to electric and $\lambda=0$ corresponds the magnetic photons, $n_{\alpha},j_{\alpha},l_{\alpha},m_{\alpha}$ is a standard set of one-electron Dirac quantum numbers. The radial integrals $\overline{M}^{(\lambda,J)}_{n_{\alpha}l_{\alpha}n_{\beta}l_{\beta}}$ are defined as
\begin{eqnarray}
\label{23}
\overline{M}^{(1,J)}_{n_{\alpha}l_{\alpha}n_{\beta}l_{\beta}}=\left[\left(\frac{J}{J+1}\right)^{1/2}\left[\left(\kappa_{\alpha}-\kappa_{\beta}\right)I^{+}_{J+1}+(J+1)I^{-}_{J+1}\right]-\left(\frac{J+1}{J}\right)^{1/2}\left[\left(\kappa_{\alpha}-\kappa_{\beta}\right)I^{+}_{J-1}-JI^{-}_{J-1}\right]\right]
\\
\nonumber
 - G\left[(2J+1)J_{J}+\left(\kappa_{\alpha}-\kappa_{\beta}\right)\left(I^{+}_{J+1}-I^{+}_{J-1}\right)-JI^{-}_{J-1}+\left(J+1\right)I^{-}_{J+1}\right],
\end{eqnarray}
\begin{eqnarray}
\label{24}
\overline{M}^{(0,J)}_{n_{\alpha}l_{\alpha}n_{\beta}l_{\beta}}=\frac{2J+1}{\left[J(J+1)\right]^{1/2}}\left(\kappa_{\alpha}+\kappa_{\beta}\right)I^{+}_{J},
\end{eqnarray}
\begin{eqnarray}
\label{25}
I^{\pm}_{J}=\int\limits_{0}^{\infty}\left(g_{\alpha}f_{\beta}\pm f_{\alpha}g_{\beta}\right)j_{J}\left(\frac{\omega r}{c}\right)dr,
\end{eqnarray}
\begin{eqnarray}
\label{26}
J_{J}=\int\limits_{0}^{\infty}\left(g_{\alpha}g_{\beta}+ f_{\alpha}f_{\beta}\right)j_{J}\left(\frac{\omega r}{c}\right)dr,
\end{eqnarray}
$g_{\alpha}$ and $f_{\alpha}$ are the large and small components of the radial Dirac wave function as defined in \cite{grant}, $\kappa$ is the Dirac angular number, $\omega$ is the photon frequency, $ j_J $ is the spherical Bessel function $G$  is the gauge parameter for the electromagnetic potentials. In our calculations we employ the "velocity" form of the matrix element containing the electromagnetic field potentials Eq. (\ref{23}) and the values $G=0$ and $\sqrt{\frac{J+1}{J}}$ \cite{labzsol}. Then we obtain what is traditionally called "velocity" and "length" gauges, respectively.

The results can be further simplified by producing explicitly the summations over magnetic quantum numbers. For this purpose we define the radial integral part for a particular combination of multipoles as
\begin{eqnarray}
\label{27}
S^{j_{n'}j_{n}}(i,j,k)=\sum_{l_{n'},l_n}\sum_{n,n'}\frac{\overline{M}^{(\lambda_i,J_{\gamma_i})}_{f,n'}(\omega_i)\overline{M}^{(\lambda_j,J_{\gamma_j})}_{n',n}(\omega_j)\overline{M}^{(\lambda_k,J_{\gamma_k})}_{n,i}(\omega_k)}{\left(E_{n'j_{n'}l_{n'}}-E_{n_fj_fl_f}-\omega_k-\omega_j\right)\left(E_{nj_nl_n}-E_{n_fj_fl_f}-\omega_k\right)}
\times
\\
\nonumber
\Delta^{j_{n'},j_{n}}\;\pi^{l_{n'}}_f(i)\;\pi^{l_n}_{n'}(j)\;\pi^{l_i}_{n_l}(k),
\end{eqnarray}
where 
\begin{eqnarray}
\pi^{l}_k(t)=
\left\{
	\begin{array}{l l}
		1\;\mbox{if}\;l_k+l+J_{\gamma_t}+\lambda_t=\mbox{odd} \\
		0\;\mbox{if}\;l_k+l+J_{\gamma_t}+\lambda_t=\mbox{even} \\ 
	\end{array}
\right. ,
\end{eqnarray}
\begin{eqnarray}
\label{28}
\Delta^{j_n,j_m}(i,j,k)=\frac{4\pi\left[j_f,j_{n'},j_n,j_i\right]^{1/2}}{\left[J_{\gamma_i},J_{\gamma_j},J_{\gamma_k}\right]^{1/2}}
\begin{pmatrix}
j_{f} & J_{\gamma_i} & j_{n'}\\
1/2 & 0 & -1/2
\end{pmatrix}
\begin{pmatrix}
j_{n'} & J_{\gamma_j} & j_{n}\\
1/2 & 0 & -1/2
\end{pmatrix}
\begin{pmatrix}
j_{n} & J_{\gamma_k} & j_{i}\\
1/2 & 0 & -1/2
\end{pmatrix}
\Theta(i,j,k)
\end{eqnarray}
and 
\begin{eqnarray}
\label{29}
\Theta(i,j,k)=\left[j_{n'},j_n\right]^{1/2}\sum_{m_{n'},m_n}(-1)^{m_i+m_f+1}
\begin{pmatrix}
j_{f} & J_{\gamma_i} & j_{n'}\\
-m_{f} & M_{\gamma_i}  & m_{n'}
\end{pmatrix}
\begin{pmatrix}
j_{n'} & J_{\gamma_j}  & j_{n}\\
-m_{n'} & M_{\gamma_j}  & m_{n}
\end{pmatrix}
\begin{pmatrix}
j_n & J_{\gamma_k}  & j_{i}\\
-m_n & M_{\gamma_k}  & m_{i}
\end{pmatrix}.
\end{eqnarray}
Here indices $i,j,k$ denotes the serial number of the photon and each one can take the values $1,2$ or $3$. The notation $[j,k, \dots ]$ means $(2j+ l)(2k+ 1)\dots$.

The final expression for the decay rate for the one-electron ion is

\begin{eqnarray}
\label{30}
\frac{dW_{i\rightarrow f}(\omega_1,\omega_2)}{d\omega_1d\omega_2} =\frac{\omega_3\;\omega_2\;\omega_1}{(2\pi)^5}\sum_{\lambda_1,\lambda_2,\lambda_3}\sum_{J_{\gamma_1}, J_{\gamma_2},J_{\gamma_3}}\sum_{M_{\gamma_1},M_{\gamma_2},M_{\gamma_3}}\left|\sum_{j_n,j_{n'}}S^{j_{n'}j_{n}}(1,2,3)+(5\;\mbox{permutations})\right|^2\;.
\end{eqnarray}
Summation over all projections appearing in the expression (\ref{30}) can be easily performed numerically for each value of corresponding angular momenta.

In the approximation of noninterecting electrons the two-electron wave function can be presented in the form
\begin{eqnarray}
\label{36}
\Psi_{J_eM_e}(\vec{r}_1,\vec{r}_2)=\frac{1}{\sqrt{2}}\sum_{m_1,m_2}C^{J_eM_e}_{j_1m_1j_2m_2}
\begin{vmatrix}
\psi_{n_1j_1l_1m_1}(\vec{r}_1)& \psi_{n_1j_1l_1m_1}(\vec{r}_2)\\\psi_{n_2j_2l_2m_2}(\vec{r}_1)&\psi_{n_2j_2l_2m_2}(\vec{r}_2)
\end{vmatrix}.
\end{eqnarray}
Then the two-electron matrix element can be reduces to the one-electron matrix element \cite{amaro}
\begin{eqnarray}
\label{37}
\left\langle n_{1_{\alpha}}n_{2_{\alpha}}J_{\alpha}M_{\alpha}\left|Q^{(\lambda)}_{J_{\gamma}M\omega}\right|n_{1_{\beta}}n_{2_{\beta}}J_{\beta}M_{\beta}\right\rangle = 
\\
\nonumber
(-1)^{J_{\alpha}-M_{\alpha}+j_{1_{\alpha}}+j_{2_{\alpha}}}\left[J_{\alpha}J_{\beta}\right]^{1/2}
\begin{pmatrix}
J_{\alpha} & J_{\gamma} & J_{\beta}\\
-M_{\alpha} & M & M_{\beta}
\end{pmatrix}
\begin{Bmatrix}
J_{\alpha} & J_{\gamma} & J_{\beta}\\
j_{1_{\alpha}} & j_2  & j_{1_{\beta}}
\end{Bmatrix}
\left\langle n_{1_{\alpha}}j_{1_{\alpha}}\left|\left|Q^{(\lambda)}_{J_{\gamma}\omega}\right|\right|n_{1_{\beta}}j_{1_{\beta}}\right\rangle ,
\end{eqnarray}
\begin{eqnarray}
\label{38}
\left\langle n_{1_{\alpha}}j_{1_{\alpha}}\left|\left|Q^{(\lambda)}_{J_{\gamma}\omega}\right|\right|n_{1_{\beta}}j_{1_{\beta}}\right\rangle
=
(-i)^{J_{\gamma}+\lambda-1}(-1)^{j_{\alpha}-1/2}\left(\dfrac{4\pi}{2J_{\gamma}+1}\right)^{1/2}\left[j_{\alpha},j_{\beta}\right]^{1/2}
\begin{pmatrix}
j_{\alpha} & J_{\gamma} & j_{\beta}\\
1/2 & 0 & -1/2
\end{pmatrix} 
\overline{M}^{(\lambda,J_{\gamma})}_{\alpha\beta}.
\end{eqnarray}
The final expression for the decay rate for two-electron ion is
\begin{eqnarray}
\label{39}
\frac{dW_{if}}{d\omega_2d\omega_1}=\frac{\omega_3\omega_2\omega_1}{(2\pi)^5}\sum_{\lambda_1,\lambda_2,\lambda_3}\sum_{J_{\gamma_1}, J_{\gamma_2},J_{\gamma_3}}\sum_{M_{\gamma_1},M_{\gamma_2},M_{\gamma_3}}\left|\sum_{J_{n'},J_n}\sum_{j_{1_{n'}},j_{1_n}}S^{J_{n'}J_nj_{1_{n'}}j_{1_n}}(1,2,3)+(5\;\mbox{permutations})\right|^2\;,
\end{eqnarray}
where
\begin{eqnarray}
\label{44}
S^{J_{n'}J_nj_{1_{n'}}j_{1_n}}(i,j,k)=\sum_{l_{n'},l_n}\sum_{n_l,n'_l}\frac{\overline{M}^{(\lambda_i,J_{\gamma_i})}_{f,n'_l}(\omega_i)\overline{M}^{(\lambda_j,J_{\gamma_j})}_{n'_l,n_l}(\omega_j)\overline{M}^{(\lambda_k,J_{\gamma_k})}_{n_l,i}(\omega_k)}{\left(E_{n'_lj_{n'}l_{n'}}-E_{n_fj_fl_f}-\omega_k-\omega_j\right)\left(E_{n_lj_nl_n}-E_{n_fj_fl_f}-\omega_k\right)}\times
\\
\nonumber
\Delta^{J_{n'},J_n,j_{1_{n'}},j_{1_n}}(i,j,k)\;\pi^{l_{n'}}_f(i)\;\pi^{l_n}_{n'_l}(j)\;\pi^{l_i}_{n_l}(k),
\end{eqnarray}
\begin{eqnarray}
\label{46}
\Delta^{J_{n'},J_n,j_{1_{n'}},j_{1_n}}(i,j,k)=\frac{(4\pi)^{3/2}\left[j_{1_f},j_{1_{n'}},j_{1_n},j_{1_i}\right]^{1/2}}{\left[J_{\gamma_i},J_{\gamma_j},J_{\gamma_k}\right]^{1/2}}
\begin{pmatrix}
j_{f} & J_{\gamma_i} & j_{n'}\\
1/2 & 0 & -1/2
\end{pmatrix}
\begin{pmatrix}
j_{n'} & J_{\gamma_j} & j_{n}\\
1/2 & 0 & -1/2
\end{pmatrix}
\begin{pmatrix}
j_{n} & J_{\gamma_k} & j_{i}\\
1/2 & 0 & -1/2
\end{pmatrix}
\\
\nonumber
\times\left[j_{1_{n'}},j_{1_n}\right]^{1/2}\left[J_{n'},J_{n}\right]^{1/2}
\begin{Bmatrix}
J_{f} & J_{\gamma_i} & J_{n'}\\
j_{1_{n'}} & j_2  & j_{1_f}
\end{Bmatrix}
\begin{Bmatrix}
J_{n'} & J_{\gamma_j} & J_{n}\\
j_{1_n} & j_2  & j_{1_{n'}}
\end{Bmatrix}
\begin{Bmatrix}
J_{n} & J_{\gamma_k} & J_{i}\\
j_{1_i} & j_2  & j_{1_n}
\end{Bmatrix}\Theta(i,j,k)
\end{eqnarray}
and
\begin{eqnarray}
\label{46a}
\Theta(i,j,k)=\left[J_{n'},J_n\right]^{1/2}\sum_{M_{n'},M_n}(-1)^{M_f+M_{n'}+M_n}
\begin{pmatrix}
J_{f} & J_{\gamma_i} & J_{n'}\\
-M_{f} & M_{\gamma_i} & M_{n'}
\end{pmatrix}
\begin{pmatrix}
J_{n'} & J_{\gamma_j} & J_{n}\\
-M_{n'} & M_{\gamma_j} & M_{n}
\end{pmatrix}
\begin{pmatrix}
J_n & J_{\gamma_k} & J_{i}\\
-M_n & M_{\gamma_k} & M_{i}
\end{pmatrix}.
\end{eqnarray}

For the intermediate states of He-like HCI in the approximation of noninteracting electrons we can use simple products of the one-electron wave functions not possessing the correct two-photon symmetry. This symmetry will be automatically restored in the two-electron matrix elements with initial or final two-electron states with proper symmetry in Eq. (\ref{36}). 

As an illustration of SSSR-2 for three equal photons we consider $2^{3}P_2\rightarrow 1^1S_0+3\gamma(E1)$ transition in the He-like HCI and demonstrate that the value $J=2$ of the total angular momentum for the three equivalent dipole photons is prohibited. In principle, this decay can proceed via several channels: 1) $2^3P_2\rightarrow n^3S_1+\gamma(E1)\rightarrow n'^3P_1\left[n'^1P_1\right]+2\gamma(E1)\rightarrow 1^1S_0+3\gamma(E1)$; this channel is prohibited since the transition $n^3S_1+\gamma(E1)\rightarrow n'^3P_1\left[n'^1P_1\right]+2\gamma(E1)$ is prohibited by SSSR-1, 2) $2^3P_2\rightarrow n^3D_1+\gamma(E1)\rightarrow n'^3P_1\left[n'^1P_1\right]+2\gamma(E1)\rightarrow 1^1S_0 +3\gamma(E1)$; this channel is also prohibited by SSSR-1 since the transition $ n^3D_1\rightarrow 1^1S_0 +2\gamma(E1)$ is prohibited by this rule, 3) $2^3P_2 \rightarrow n^3D_2\left[n^1D_2\right]+\gamma(E1)\rightarrow n'^3P_1\left[n'^1P_1\right]+2\gamma(E1)\rightarrow 1^1S_0$. The states admixed by the spin-orbit interaction are placed in the square brackets, $n,\, n'$ are arbitrary integer numbers. The contribution of the channel 3) does not turn to zero so evidently. 

However, using an exact expression (\ref{39}) we find that for the three equal-frequency photons ($\omega_1=\omega_2=\omega_3$) the total expression for the transition rate turns to zero. This can be traced analytically from Eq. (\ref{39}), but we also illustrate it by numerical calculations for $Z=92$ (see Fig. 3). Note that in the channel 1) there is the resonance (the situation when the energy denominator turns to zero), in this case: $2^{3}P_2\rightarrow 2^3S_1+\gamma(E1)\rightarrow 1^1S_0+3\gamma(E1)$. The presence of the cascade-producing state $2^3S_1$ in the sum over all the intermediate states in the transition amplitude Eq. (\ref{44}) leads to the arrival of the high, but narrow "ridge" in the frequency distribution $\frac{dW(\omega_1,\omega_2)}{d\omega_1d\omega_2}$ starting at the point $\omega_1=E(2^3P_2)-E(2^3S_1)$ on the $\omega_1$ axis and extending parallel to the $\omega_2$ axis; alternatively, this "ridge" can start from the point $\omega_2=E(2^3P_2)-E(2^3S_1)$ and extend parallel to the $\omega_1$ axis. The "ridge" does not influence the existence of the SSSR-2: it does not correspond to the case of three equivalent photons. This is a general conclusion for all the possible cascade transitions.
 In Fig. 3 one can see "pit" in the frequency distribution for the transition rate which arises around the point ($\omega_1=\omega_2=\omega_3$) due to the SSSR-2. For comparison in Fig. 4 we present a picture for the frequency distribution for transition $2^3P_1\rightarrow 1^1S_0+3\gamma(E1)$ also for $Z=92$, where instead of the SSSR-2-induced "pit" there is a "hump", corresponding to the maximum in the two-dimensional distribution Fig. 1 (left panel). In this transition the total angular momentum $J=1$ is not prohibited for 3 equivalent dipole photons.

In the same way (using Eq. (\ref{39})) we prove SSSR-2 in case of transition $3^3D_2\rightarrow 1^1S_0+3\gamma(M1)$. However, neither the value $J=2$ for the total angular momentum of 3-photon system is forbidden for the transition $3^3D_2\rightarrow 1^1S_0+3\gamma(E2)$, nor the value $J=4$ is forbidden for $4^3F_4\rightarrow 1^1S_0+3\gamma(M2)$ transition. The same concerns $2^1S_0\rightarrow 3\gamma(E2)$ transition where the value $J=0$ is not forbidden for the 3-photon system. In all these cases the value $J$ is fixed by the initial $J_{e_i}$ and final $J_{e_f}$ values for the total electronic angular momentum. All these examples confirm our remark in the Introduction that the SSSR-2 works, in general, only for the dipole photons. 

In some cases, for example for transition  $2^1S_0\rightarrow 1^1S_0+3\gamma(E1)$ the $J=0$ value is excluded not by the SSSR-2 but for more trivial reasons: according to SSSR-1 the two $E1$ photons can have a total angular momentum $0,2$. Adding the angular momentum $1$ of the third $E1$ photon to these values, it is impossible to receive the total angular momentum $J=0$ for 3-photon system. With the three $E2$  photons the situation is different: according to SSSR-1 a two-photon system can have values of the total angular momentum equal to $0,2,4$. Adding to these values the value $2$ for the angular momentum of the third $E2$ photon one can receive the value $J=0$ for the total angular momentum of three photons. A direct check with Eq. (\ref{39}) shows that the value $J=0$, is not forbidden in this case.

\section{General analytic proof of the SSSR-2}

In this section we give a general proof of the SSSR-2 not connected with any particular transitions in atoms. For this purpose we will use the equations for the Coefficients of Fractional Parentage (CFP) usually employed in case of Fermi-Dirac statistics for the construction of the wave functions for the groups of the equivalent electrons (see, for example, \cite{sobelman}).

We consider a system of 3 photons with equal frequencies $\omega$, equal angular momenta $j_1=j_2=j_3=j$ and of the same type: $s_1=s_2=s_3=s$. At first we will distinguish these quantum numbers and will put them equal to each other later. We choose one possible coupling scheme for three photons where at first we couple the photons with indices $i=1,2$ and then add to them the third photon with index $i=3$. The wave function for the pair of photons with $i=1,2$ we construct according to Eq. (\ref{8}) and then add the third photon $i=3$ using only the standard angular momentum coupling scheme and aiming to obtain the 3-photon wave function with the total angular momentum $J$ and its projection $M$. We denote this wave function as $\Phi_{\omega JM}(j_1j_2[J']j_3)$, where the angular momentum $J'$ in square brackets denotes one of the possible values for the intermediate total angular momentum for two photons. The wave function $\Phi_{\omega JM}(j_1j_2[J']j_3)$ is symmetrized with respect to the permutations of the photons $1,2$ and therefore $J'$ takes only even values according to SSSR-1. Then we decompose this wave functions via the wave functions corresponding to another coupling scheme $\Phi_{\omega JM}(j_2j_3[J']j_1)$. The latter wave functions are not yet symmetrized with respect to the photon permutations $2,3$. The decomposition looks like
\begin{eqnarray}
\label{5.1}
\Phi_{\omega JM}(j_1 j_2 [J'] j_3)=\sum\limits_{J''}\left(j_2 j_3 [J''] j_1 J | j_1 j_2 [J'] j_3 J\right) \Phi_{\omega JM}(j_2j_3[J'']j_1),
\end{eqnarray}
where $\left(j_2 j_3 [J''] j_1 J | j_1 j_2 [J'] j_3 J\right)$ are the Racah coefficients connected with $6j$-symbols via
\begin{eqnarray}
\label{5.2}
\left(j_2 j_3 [J''] j_1 J | j_1 j_2 [J'] j_3 J\right) = (-1)^{j_1+j_2+j_3+J}\sqrt{(2J'+1)(2J''+1)}
\begin{Bmatrix}
j_1 & j_2 & J'\\
j_3 & J  & J''
\end{Bmatrix}.
\end{eqnarray}

The summation in Eq. (\ref{5.1}) extends over all the values $J''$ allowed by the vector coupling scheme for two angular momenta $j_2$, $j_3$. To ensure the symmetry of the wave function with respect to the permutation of the variables $2,3$, according to SSSR-1 we have to retain only the even values of $J''$ in these expansion. For this purpose we replace the wave function Eq. (\ref{5.1}) by the linear combination
\begin{eqnarray}
\label{5.3}
\Phi_{\omega JM}(j_1 j_2 j_3)=\sum\limits_{J'}\left(j_1 j_2 [J'] j_3 J \} j_1 j_2 j_3 J \right)\Phi_{\omega JM}(j_1 j_2 [J'] j_3),
\end{eqnarray}
where $\left(j_1 j_2 [J'] j_3 J \} j_1 j_2 j_3 J \right)$ are the CFP and define the CFP from the requirement that all the terms with odd values of $J''$ in the expansion
\begin{eqnarray}
\label{5.4}
\Phi_{\omega JM}(j_1 j_2 j_3)=\sum\limits_{J'}\left(j_1 j_2 [J'] j_3 J \} j_1 j_2 j_3 J \right)\sum\limits_{J''}\left(j_2 j_3 [J''] j_1 J | j_1 j_2 [J'] j_3 J \right)\Phi_{\omega JM}(j_2j_3[J'']j_1)
\end{eqnarray}
turn to zero. Since the presence of the odd $J''$ values in the wave function Eq. (\ref{5.1}) contradicts to the requirement of its symmetrization with respect to the permutation of variables $2,3$, the absence of these terms guarantees the corresponding symmetry. It is easy to check that any function of three variables $1,2,3$, symmetric with respect to permutations $1,2$ and $2,3$ will be symmetric also with respect to the permutation $1,3$. Thus the absence of the odd $J''$ values in the expansion Eq. (\ref{5.4}) makes the wave function $\Phi_{\omega JM}(j_1 j_2 j_3)$ fully symmetric with respect to the permutations of the photon variables.

The condition for the absence of the odd $J''$ values in the expansion Eq. (\ref{5.4}) reduces to the system of equations
\begin{eqnarray}
\label{5.5}
\sum\limits_{J'}\left(j_1 j_2 [J'] j_3 J \} j_1 j_2 j_3 J \right)\left(j_2 j_3 [J''] j_1 J | j_1 j_2 [J'] j_3 J \right)=0
\end{eqnarray}
for all possible odd $J''$ values. The summation in Eq. (\ref{5.5}) extends over all possible even $J'$ values for the two-photon system. Setting now $j_1=j_2=j_3=j$ we rewrite this equations as
\begin{eqnarray}
\label{5.6}
\sum\limits_{J'}\left(j j [J'] j J \} j j j J \right)\left(j j [J''] j J | j j [J'] j J \right)=0.
\end{eqnarray}
For the CFP in case of equivalent photons we will use shorthand notation
\begin{eqnarray}
\label{5.7}
\left(j j [J'] j J \} j j j J \right)\equiv C^j_{J'\,J}.
\end{eqnarray}
Then with Eq. (\ref{5.2}) taken into account for the equivalent photons Eq. (\ref{5.6}) reduces to 
\begin{eqnarray}
\label{5.8}
\sum\limits_{J'} C^j_{J'\,J} \sqrt{2J'+1} 
\begin{Bmatrix}
j & j & J'\\
j & J  & J''
\end{Bmatrix}
=0.
\end{eqnarray}
For the dipole photons we have to set $j=1$ and according to SSSR-1 $J'=0,2$, $J''=1$. This yields
\begin{eqnarray}
\label{5.9}
C^1_{0\, J}
\begin{Bmatrix}
1 & 1 & 0\\
1 & J  & 1
\end{Bmatrix}
+
C^1_{2\, J}\sqrt{5}
\begin{Bmatrix}
1 & 1 & 2\\
1 & J  & 1
\end{Bmatrix}
=0.
\end{eqnarray}

In this case we have 1 equation for two coefficients $C^1_{0\, J}$ and $C^1_{2\, J}$. In principle, normalization condition for the function Eq. (\ref{5.3}) would give one more equation but it is not necessary to employ it for our purposes. The first $6j$-symbol in Eq. (\ref{5.9}) is equal to
\begin{eqnarray}
\label{5.10}
\begin{Bmatrix}
1 & 1 & 0\\
1 & J  & 1
\end{Bmatrix}
=(-1)^J\frac{1}{\sqrt{3(2J+1)}}\delta_{J1}.
\end{eqnarray}
According to the angular momentum coupling rules the possible $J$ values for 3 dipole photons are $J=1,2,3$. If we set $J=1$ Eq. (\ref{5.9}) always has a solution with nonzero values of $C^1_{0\, 1}$, $C^1_{2\, 1}$. However, setting $J=2$ we arrive at the equality
\begin{eqnarray}
\label{5.11}
C^2_{2\, 2}\sqrt{5}
\begin{Bmatrix}
1 & 1 & 2\\
1 & 2  & 1
\end{Bmatrix}
=0.
\end{eqnarray}
Since $6j$-symbol in Eq. (\ref{5.11}) is not equal to zero, it follows that $G^1_{2\, 2}=0$. This means that for the $J=2$ value of the total angular momentum for three dipole photons in the wave function Eq. (\ref{5.4}) the value $J'=2$ for the two-photon subsystem should be absent. This contradicts to the SSSR-1 and we conclude that the existence of the total angular momentum value $J=2$ for the three-photon system is inconsistent with the SSSR-1 for the two-photon system. Thus the value $J=2$ is forbidden for the system of three equivalent dipole photons. It remains to consider the case $J=3$. For $J=3$ Eq. (\ref{5.9}) reads
\begin{eqnarray}
\label{5.12}
C^1_{2\, 3}\sqrt{5}
\begin{Bmatrix}
1 & 1 & 2\\
1 & 3  & 1
\end{Bmatrix}
=0.
\end{eqnarray}
However,
\begin{eqnarray}
\label{5.13}
\begin{Bmatrix}
1 & 1 & 2\\
1 & 3  & 1
\end{Bmatrix}
=0
\end{eqnarray}
and the coefficient $C^1_{2\, 3}$ is nonzero, but arbitrary. This does not contradict to the SSSR-1 and the vallue $J=3$ for the total angular momentum of the three equivalent dipole photon system is allowed. This proves the SSSR-2.

\section{SSSR-3 for 4 dipole photon transitions}

It is important to check SSSR for the 4-photon transitions and especially for the higher $J$ values since the fermionic analogue for the SSSR is violated as we have seen in section III, for 4-electron systems and $J_e=5$. Therefore, we give first the analytic proof of the SSSR-3 for 4 dipole photons.

Proceeding in the same way as for the proof of SSSR-2, we consider two coupling schemes for constructing of the wave function of the 4-photon system. The first scheme corresponds to the wave function
\begin{eqnarray}
\label{5.14}
\Phi_{\omega JM}\left(j_1 j_2 [J_{12}] j_3 j_4 [J_{34}] JM\right),
\end{eqnarray}
where we first couple momenta $j_1 j_2$ to the intermediate momentum $J_{12}$, then couple momenta $j_3 j_4$ to another intermediate momentum $J_{34}$ and finally couple two intermediate momenta $J_{12}$ and $J_{34}$ to the final total angular momentum of the 4-photon system $J$ with the projection $M$. We assume that the wave function Eq. (\ref{5.4}) is symmetric with respect to the permutation of the variables $1,2$ and with respect to the permutation of variables $3,4$. Then according to the SSSR-1 the momenta $J_{12}$ and $J_{34}$ can take only even values. Another coupling scheme will be representeed by the wave function
\begin{eqnarray}
\label{5.15}
\Phi_{\omega JM}\left(j_1 j_3 [J_{13}] j_2 j_4 [J_{24}] JM\right).
\end{eqnarray}

In this wave function we do not assume the symmetrization with respect to variables $1,3$ and $2,4$ and decompose the wave function Eq. (\ref{5.14}) in terms of the wave functions Eq. (\ref{5.15}):
\begin{eqnarray}
\label{5.16}
\Phi_{\omega JM}\left(j_1 j_2 [J_{12}] j_3 j_4 [J_{34}] JM\right)=\sum\limits_{J_{13}J_{24}}\left(j_1 j_3 [J_{13}] j_2 j_4 [J_{24}] J | j_1 j_2 [J_{12}] j_3 j_4 [J_{34}] J \right)
\Phi_{\omega JM}\left(j_1 j_3 [J_{13}] j_2 j_4 [J_{24}] JM\right),
\end{eqnarray}
where $\left(j_1 j_3 [J_{13}] j_2 j_4 [J_{24}] J | j_1 j_2 [J_{12}] j_3 j_4 [J_{34}] J \right)$ are the Fano coefficients connected with $9j$-symbols via
\begin{eqnarray}
\label{5.17}
\left(j_1 j_2 [J_{12}] j_3 j_4 [J_{34}] J | j_1 j_3 [J_{13}] j_2 j_4 [J_{24}] J \right)=
\\
\nonumber
\sqrt{(2J_{12}+1)(2J_{34}+1)(2J_{13}+1)(2J_{24}+1)}
\begin{Bmatrix}
j_1 & j_2 & J_{12}\\
j_3 & j_4 & J_{34}\\
J_{13} & J_{24} & J
\end{Bmatrix}.
\end{eqnarray}

In Eq. (\ref{5.16}) in the expansion over $J_{13}$, $J_{24}$ all the values of $J_{13}$ $J_{24}$ allowed by the angular momenta vector coupling are present. To symmetrize the wave function Eq. (\ref{5.16}) with respect to the permutation of variables $1,3$ and with respect to the permutation $2,4$ we have to replace it by the linear combination
\begin{eqnarray}
\label{5.18}
\Phi_{\omega JM}\left(j_1 j_2 j_3 j_4\right)=\sum\limits_{J_{12}J_{34}}
\left(j_1 j_2 [J_{12}] j_3 j_4 [J_{34}] J \} j_1 j_2 j_3 j_4 J \right)\Phi_{\omega JM}\left(j_1 j_2 [J_{12}] j_3 j_4 [J_{34}] J M\right) = 
\\
\nonumber
\sum\limits_{J_{12}J_{34}}
\sum\limits_{J_{12}J_{34}}
\left(j_1 j_2 [J_{12}] j_3 j_4 [J_{34}] J \} j_1 j_2 j_3 j_4 J \right)
\left(j_1 j_3 [J_{13}] j_2 j_4 [J_{24}] J \} j_1 j_2 [J_{12}] j_3 j_4 [J_{34}] J \right)\times
\\
\nonumber
\Phi_{\omega JM}\left(j_1 j_3 [J_{13}] j_3 j_4 [J_{24}] J M\right),
\end{eqnarray}
where $\left(j_1 j_2 [J_{12}] j_3 j_4 [J_{34}] J \} j_1 j_2 j_3 j_4 J \right)$ are the CFP for the 4-particle systems (bosons). For symmetrization of the wave function Eq. (\ref{5.18}) with respect to the permutation of variables $1,3$ and with respect to the permutation $2,4$ it is necessary to require the terms with odd values of $J_{13}$, $J_{24}$ to vanish in the summation over $J_{13}$, $J_{24}$ in Eq. (\ref{5.18}). This requirement leads to the system of equations for the CFP:
\begin{eqnarray}
\label{5.19}
\sum\limits_{J_{13} J_{24}}\left(j_1 j_2 [J_{12}] j_3 j_4 [J_{34}] J\} j_1 j_2 j_3 j_4 J\right)
\left(j_1 j_3 [J_{13}] j_2 j_4 [J_{24}] J | j_1 j_2 [J_{12}] j_3 j_4 [J_{34}] J \right)=0,
\end{eqnarray}
where $J_{13}$, $J_{24}$ take only the odd values and the summation is extended over the even values of $J_{13}$, $J_{24}$. When these equations are satisfied we can consider the wave function Eq. (\ref{5.18}) as symmetric with respect to the permutations of the variables $1,3$ with each other and the variables $2,4$ with each other. If any function of 4 variables is symmetric with respect to the permutations within pairs $(1,2)$, $(3,4)$, $(1,3)$ and $(2,4)$ it is fully symmetric. Indeed, let us fix for example, variable 4 and consider the permutations within a group of variables $1,2,3$. If the function is symmetric with respect to the permutation $1,2$ and to the permutation $1,3$ it will be symmetric also with respect to the  permutation $2,3$ as it follows from the symmetrization of the functions of the three variables. In this way we can prove the symmetry with respect to arbitrary pair of variables. Thus the wave function of the four photon constructed as described above will be fully symmetric, i.e. will obey the Bose-Einstein statistics.

Remembering that we consider equivalent photons $j_1=j_2=j_3=j_4=j$ and using the shorthand notations for the 4-particle CFP
\begin{eqnarray}
\label{5.20}
\left(j j [J_{12}] j j [J_{34}] J\} j j j j J\right)\equiv G^j_{J_{12},\, J_{34},\, J}
\end{eqnarray}
we rewrite Eq. (\ref{5.19}) as
\begin{eqnarray}
\label{5.21}
\sum\limits_{J_{12}, J_{34}}G^j_{J_{12},\, J_{34},\, J}\sqrt{(2J_{12}+1)(2J_{34}+1)}
\begin{Bmatrix}
j & j & J_{12}\\
j & j & J_{34}\\
J_{13} & J_{24} & J
\end{Bmatrix}=0.
\end{eqnarray}
Going over to the case of dipole photons $(j=1)$ we should extend the summation in Eq. (\ref{5.21}) over the values $J_{12}=0,2$ and $J_{34}=0,2$. In the right-hand side of Eq. (\ref{5.21}) we have to set $J_{13}=1$, $J_{24}=0,2$ or $J_{13}=0,2$, $J_{24}=1$ or $J_{13}=J_{24}=1$. All these cases should be excluded from the summation over $J_{13}$, $J_{24}$ in Eq. (\ref{5.18}). Eq. (\ref{5.21}) now looks like
\begin{eqnarray}
\label{5.22}
\sum\limits_{J_{12}, J_{34}}G^1_{J_{12},\, J_{34},\, J}\sqrt{(2J_{12}+1)(2J_{34}+1)}
\begin{Bmatrix}
1 & 1 & J_{12}\\
1 & 1 & J_{34}\\
J_{13} & J_{24} & J
\end{Bmatrix}=0.
\end{eqnarray}

In general, for the fixed $J$ value there are 5 equations for the 4 coefficients $G^1_{0,\, 0,\, J}$, $G^1_{2,\, 0,\, J}$, $G^1_{0,\, 2,\, J}$ and $G^1_{2,\, 2,\, J}$. However, two of these equations for the case $J_{13}=1$, $J_{24}=0,2$ coincide with the other two equations for the case $J_{13}=0,2$, $J_{24}=1$ due to invariance of $9j$-symbol in Eq. (\ref{5.21}) under the permutation of the two first rows. Then the system of equations looks like
\begin{eqnarray}
\label{5.23}
G^1_{0,\, 0,\, J}
\begin{Bmatrix}
1 & 1 & 0\\
1 & 1 & 0\\
1 & 0 & J
\end{Bmatrix}+
G^1_{0,\, 2,\, J}
\begin{Bmatrix}
1 & 1 & 0\\
1 & 1 & 2\\
1 & 0 & J
\end{Bmatrix}\sqrt{5}+
G^1_{2,\, 0,\, J}
\begin{Bmatrix}
1 & 1 & 2\\
1 & 1 & 0\\
1 & 0 & J
\end{Bmatrix}\sqrt{5}+
G^1_{2,\, 2,\, J}
\begin{Bmatrix}
1 & 1 & 2\\
1 & 1 & 2\\
1 & 0 & J
\end{Bmatrix}5=0,
\end{eqnarray}
\begin{eqnarray}
\label{5.24}
G^1_{0,\, 0,\, J}
\begin{Bmatrix}
1 & 1 & 0\\
1 & 1 & 0\\
1 & 2 & J
\end{Bmatrix}+
G^1_{0,\, 2,\, J}
\begin{Bmatrix}
1 & 1 & 0\\
1 & 1 & 2\\
1 & 2 & J
\end{Bmatrix}\sqrt{5}+
G^1_{2,\, 0,\, J}
\begin{Bmatrix}
1 & 1 & 2\\
1 & 1 & 0\\
1 & 2 & J
\end{Bmatrix}\sqrt{5}+
G^1_{2,\, 2,\, J}
\begin{Bmatrix}
1 & 1 & 2\\
1 & 1 & 2\\
1 & 2 & J
\end{Bmatrix}5=0,
\end{eqnarray}
\begin{eqnarray}
\label{5.25}
G^1_{0,\, 0,\, J}
\begin{Bmatrix}
1 & 1 & 0\\
1 & 1 & 0\\
1 & 1 & J
\end{Bmatrix}+
G^1_{0,\, 2,\, J}
\begin{Bmatrix}
1 & 1 & 0\\
1 & 1 & 2\\
1 & 1 & J
\end{Bmatrix}\sqrt{5}+
G^1_{2,\, 0,\, J}
\begin{Bmatrix}
1 & 1 & 2\\
1 & 1 & 0\\
1 & 1 & J
\end{Bmatrix}\sqrt{5}+
G^1_{2,\, 2,\, J}
\begin{Bmatrix}
1 & 1 & 2\\
1 & 1 & 2\\
1 & 1 & J
\end{Bmatrix}5=0.
\end{eqnarray}
Eq. (\ref{5.23}) corresponds to the case $J_{13}=1$, $J_{24}=0$, Eq. (\ref{5.24}) corresponds to the case $J_{13}=1$, $J_{24}=2$ and Eq. (\ref{5.25}) corresponds to the case $J_{13}=J_{24}=1$.

First, we analyze the system of equations Eq. (\ref{5.23})-Eq. (\ref{5.25}) for $J=1$. Considering Eq. (\ref{5.23}) we see that all the $9j$-symbols in the left-hand side of Eq. (\ref{5.23}) but the last one are zero \cite{Varsh}. The last $9j$-symbol is nonzero only for $J=1$:
\begin{eqnarray}
\label{5.26}
\begin{Bmatrix}
1 & 1 & 2\\
1 & 1 & 2\\
1 & 0 & J
\end{Bmatrix}
=-\frac{1}{\sqrt{15}}
\begin{Bmatrix}
1 & 2 & 1\\
2 & 1 & 1
\end{Bmatrix}
\delta_{J\, 1}.
\end{eqnarray}
This means that for $J=1$ $G^1_{2,\, 2,\, 1}=0$ and the value $J=1$ for the total angular momentum of 4-photon system is inconsistent with SSSR-1 which allows the value $J=2$ for the two-electron system (condition $G^1_{2,\, 2,\, 1}=0$ means that the value $J=2$ is not allowed for the two-photon subsystems of the 4-photon system). All other possible values $J=0,2,3,4$ for the 4-dipole-photon system are not forbidden by Eq. (\ref{5.23}). Continuing this analysis we find that in Eq. (\ref{5.24}) for $J=0$ all the $9j$-symbols turn to zero, so this equation has a solution with arbitrary values of the CFP and the value $J=0$ is not forbidden for the 4-equal-dipole-photon system. For $J=1$ in Eq. (\ref{5.24}) all the $9j$-symbols but the last one are zero. Then $G^1_{2,\, 2,\, 1}=0$ what contradicts to SSSR-1 and the value $J=1$ is forbidden. For $J=2$ only the first $9j$-symbol in Eq. (\ref{5.24}) is zero, so that Eq. (\ref{5.24}) allows for the nonzero solution for CFPs and the value $J=2$ is allowed. For $J=3$ all the $9j$-symbols but the last one are zero, so $G^1_{2,\, 2,\, 3}=0$ which contradicts again to the SSSR-1 and the value $J=3$ is forbidden. In the same way we will find that Eq. (\ref{5.25}) forbids only the value $J=1$. Finally, for $J=4$ all the $9j$-symbols in Eq. (\ref{5.24}) (as well as in Eqs. (\ref{5.23}), (\ref{5.25})) are zero. Then the CFP with $J=4$ are fully arbitrary and the value $J=4$ is allowed for the 4-photon system. In total, our analysis demonstrates that for the 4-equal-dipole-photon system the values of the total angular momentum of the system $J=0,2,4$ are allowed and the values $J=1,3$ are forbidden.

To support the general proof and using the same QED approach as for 3-photon systems we have checked the SSSR-3 for the transitions $2^3S_1\rightarrow 1^1S_0+4\gamma(E1)$ and $3^3D_3\rightarrow 1^1S_0+4\gamma(E1)$. In both cases transition rates turn to zero for equal frequency photons $\omega_1=\omega_2=\omega_3=\omega_4$. This can be traced analytically from equations similar to Eq. (\ref{39}) and proves that the value $J=1$ is forbidden for the equal-frequency photons in the first transition and $J=3$ is forbidden in the second transition. Thus, SSSR-3 holds for the 4-photon transitions.



\section{conclusions and outlook}

We have formulated the Spin-Statistic Selection Rules (SSSR) for the multiphoton atomic processes which originate from the fundamental properties of spin-1 particles, obeying the Bose-Einstein statistics. In a sense, these rules can be considered as an exclusion principle for bosons (photons) since they prohibit some states for the system of equivalent particles. This resemblance is strengthened by comparison with the properties of the equivalent electrons in atomic physics. 

However the SSSR are formulated exclusively for the atomic processes, i.e. for the emission or absorption of photons by atomic systems (atoms, ions). The SSSR are related to the total angular momentum quantum numbers, the orbital angular momenta of photons also being included. The latter ones depend on the choice of the frame of reference, in case of SSSR this choice is the rest frame of an atom emitting or absorbing the photons.

This makes the difference with the Landau-Yang theorem which states that two-photon system can not have a total angular momentum equal to one. This statement is formulated in the rest frame for the center-of-inertia for two photons. In this reference frame two photons are collinear (opposite directed) and have equal frequencies. The different choice of the reference frame compared to the SSSR leads to a different formulations when the higher multipolarities are involved , i.e. the orbital angular momentum definition begins to be important. Otherwise the SSSR can be considered as an extension of the Landau-Yang theorem to multiphoton systems in atomic processes.

Our calculations were performed for the two-electron highly charged ions. That is, they were fully relativistic but with full neglect of the interelectron interaction. We have chosen the He-like HCI because with the neglect of the interelectron interaction the calculations are as simple as for the one-electron ions. On the other side the He-like ions have essentially more reach spectrum which allows to find many possibilities for the application of the SSSR. The SSSR, as based on the symmetry properties remain the same independent on the inclusion or neglect of the interelectron interaction . Of course the numerical values for the transition rates obtained in such an approximation will be far from accurate for the neutral helium, but will become more accurate for the HCI with high $Z$ values. This is the reason why we have chosen uranium ($Z=92$) for our particular examples. The SSSR should hold not only for two-electron atoms and ions but also for the multiphoton processes in any many-electron atoms or ions. Our choice of the two-electron HCI was made because in these systems the action of the SSSR becomes most transparent.

An important question is in which kind of the experiments the influence of the SSSR on atomic processes can be observed. It is natural to use lasers for such observation. Let the laser light propagate through atomic vapour of an atom with suitable transition frequency $\omega_a$  between the pair of atomic levels. An advantage of the use of the laser source is that all the photons will have the same frequency. If we divide this frequency by an integer number $N_{\gamma}$ and adjust the laser frequency $\omega_l$ to this value, $\omega_l=\omega_a/N_{\gamma}$, the number of photons $N_{\gamma}$ in the absorption process will be fixed. The value of the total angular momentum $J$ for N-photon system can be fixed by choosing the appropriate values $J_{e_i}$ and $J_{e_f}$ for the initial (lower) and final (upper) levels in the transition process. For example, if we choose $J_{e_i}=0$, $J_{e_f}=2$ and $N_{\gamma}=3$ we will check the SSSR-2 for $N_{\gamma}=3, J=2$. What we can not  fix is the multipolarity of the transition, i.e. a total angular momentum $j$ of every separate photon. A laser light in the beam can be decomposed in all possible multipolarities. This means, for example that together with $E1E1E1$ transition, all transitions with the same total parity constructed with the higher multipoles, i.e. $E1M1E2$, $E1E1M2$ etc. will be always absorbed. However the process with the photons of higher multipolarities are usually strongly suppressed in atoms. Due to this suppression the $E1E1E1$ transition will be dominant. Measuring the absorption rate at the $\omega_l=\omega_a/N_{\gamma}$ frequency one can establish the validity or non-validity of the particular SSSR.


\begin{center}
Acknowledgments
\end{center}
The work was supported by RFBR (grants No. 12-02-31010 and No. 14-02-00188). T. Z., D. S. and L. L. acknowledge the support by St.-Petersburg State University with a research grant 11.38.227.2014. D. S. is grateful to the Max Planck Institute for the Physics of Complex Systems for financial support and to the Dresden University of Technology for hospitality.

\begin{figure}[hbtp]
\caption{Photon frequency distributions for the two-photon transitions $ (1s2s)^1S_0\rightarrow (1s)^{2\;1}S_0+2\gamma (E1) $ (left panel) and $ (1s2s)^{3}S_1\rightarrow (1s)^{2\;1}S_0+2\gamma (E1)$ (right panel)  in He-like U (nuclear charge $ Z=92 $). The transition rates $ \frac{dW}{d\omega} $ in units $ s^{-1} $ is plotted versus the photon frequency $ \omega $ in units $ \omega/\Delta_{0} $ and $ \omega/\Delta_{1} $. The values $ \Delta_{0} $, $ \Delta_{1} $ correspond to the energy intervals $ \Delta_{0}=E(2^1S_0)-E(1^1S_0) $ and $ \Delta_{1}=E(2^3S_1)-E(1^1S_0) $, respectively. The calculations are performed fully relativistically, with Dirac one-electron wave functions and relativistic expressions for the electromagnetic vector potentials (photon wave functions). The Dirac wave functions for an extended nucleus described by the Fermi distribution were employed. The interelectron interaction was fully neglected; the expected error is about $ 1/Z $ i.e. about $ 1\%$ for $Z=92$. The one-electron wave functions for the initial and final two-electron states were coupled to present $ 1^1S_0,2^1S_0 $ and $ 2^3S_1 $ states, respectively. Summation over the full set of two-electron states taken as products of one-electron Dirac states was performed within the B-spline approach. The calculations were carried out in two relativistic "forms", corresponding to the nonrelativistic "length" and "velocity" forms \cite{labzsol}; the results coincide with 12 digits. For the ions with lower $ Z $ values and for the neutral He atom the accuracy for $ W $ values obtained from calculations with the neglect of interelectron interaction becomes poorer, but the difference between the frequency distributions for $ 2^1S_0\rightarrow 1^1S_0+2\gamma(E1) $ and $ 2^3S_1\rightarrow 1^1S_0+2\gamma(E1) $ transitions  qualitatively remains the same.}
\includegraphics[scale=0.97]{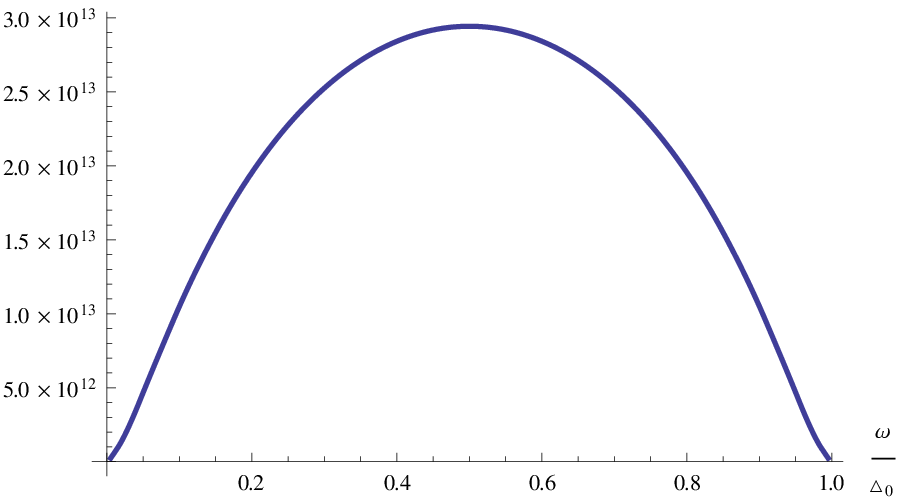}
\includegraphics[scale=0.97]{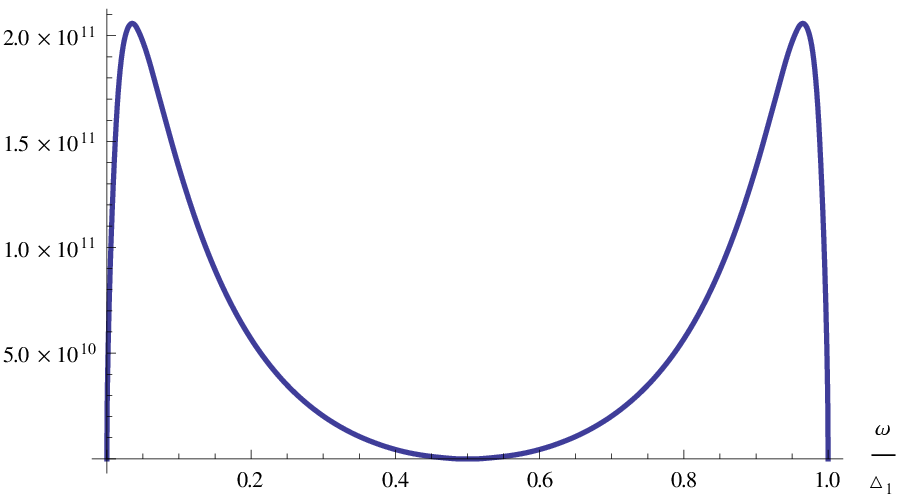}
\end{figure}

\begin{figure}[hbtp]
\caption{Photon frequency distribution for the $(1s3d)^3D_3\rightarrow (1s)^{2;1}S_0+2\gamma(E2)$ two-photon transition. Notations, units etc. are the same as in Fig. 1. $\Delta_{3}$ denotes the energy difference $\Delta_{3} = E(3^3D_3)-E(1^1S_0)$. Calculation is performed for He-like U ion ($Z=92$). The total transition probability (integral value) is $W_{2\gamma}(3^3D_3-1^1S_0)=3.3299\times10^4$ $s^{-1}$, $W_{2\gamma}(3^3D_3-1^1S_0)=3.3299\times10^4$ $s^{-1}$ in the relativistic "length" and "velocity" forms respectively.}
\centering
\includegraphics[scale=0.97]{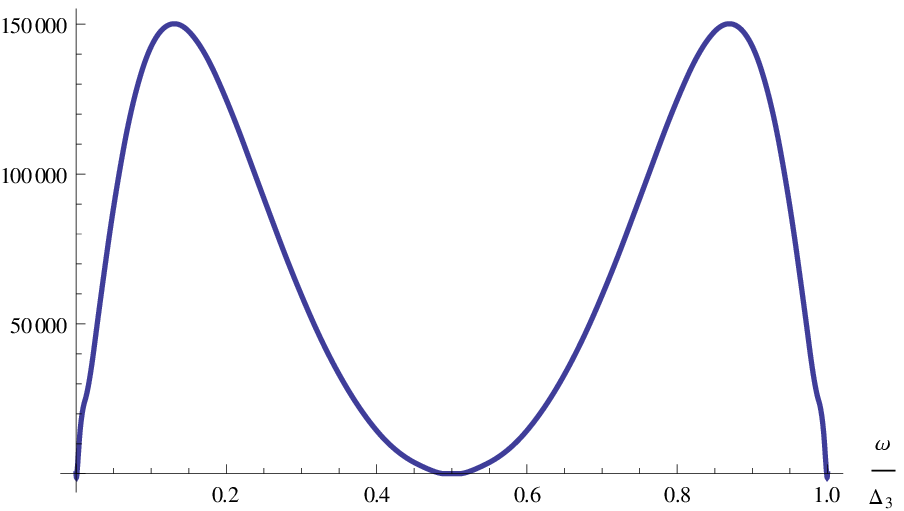}
\end{figure}

\begin{figure}[hbtp]
\caption{3-dimensional plot for frequencies distribution of the transition rate $ 2^3P_2\rightarrow 1^1S_0+3\gamma(E1) $ in He-like uranium. On the vertical axis the transition rate $ \frac{dW}{d\omega_1d\omega_2}$ in $ s^{-1} $ is plotted; on the horizontal axes the photon frequencies are plotted in units $ \omega_1/\Delta_2 $, $ \omega_2/\Delta_2  $ where  $ \Delta_{2} $ denotes the energy difference $ \Delta_{2}=E(2^3P_2)-E(1^1S_0) $. The lowest (zero) point is the point with coordinates $ \omega_1/\Delta_2=\omega_2/\Delta_2=1/3 $ at the bottom of the "pit" in the frequency distribution for the transition rate which arises due to SSSR-2. The calculations were done neglecting the interelectron interaction.}
\centering
\includegraphics[scale=0.97]{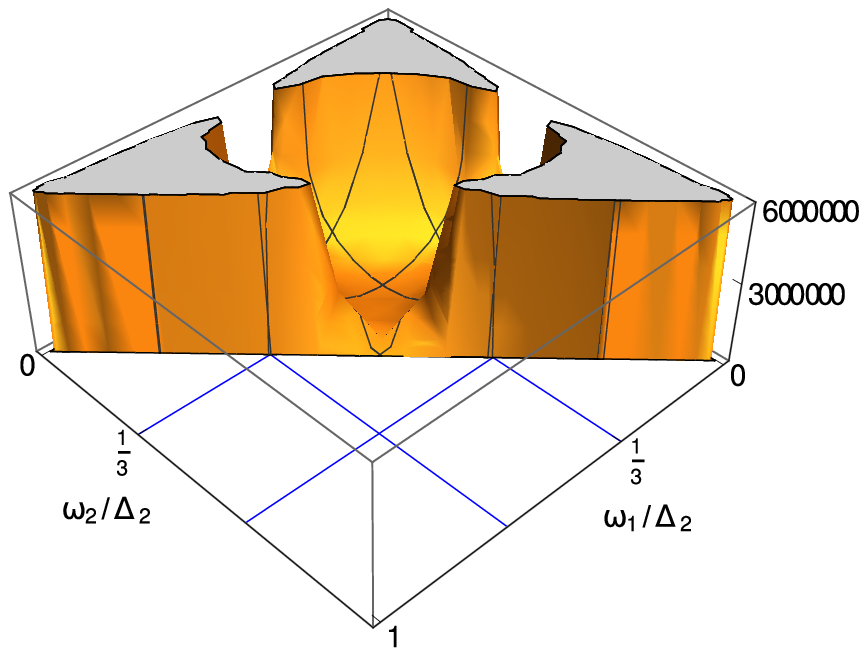}
\end{figure}

\begin{figure}[hbtp]
\caption{3-dimensional plot for frequencies distribution of the transition rate $ 2^3P_1\rightarrow 1^1S_0 +3\gamma(E1) $ in He-like uranium. All the details are the same as in Fig. 3. $ \Delta_{1}=E(2^3P_1)-E(1^1S_0) $. In this case, there is no SSSR-2 induced "pit" in the frequency distribution for the transition rate. The calculations are performed under the same approximation as for Fig. 3. The total transition probability (integral value) is $W_{3\gamma}(2^3P_1-1^1S_0)=13.825270\times 10^6$ $s^{-1}$, $W_{3\gamma}(2^3P_1-1^1S_0)=13.825246\times 10^6$ $s^{-1}$ in the relativistic "length" and "velocity" forms respectively.}
\centering
\includegraphics[scale=0.97]{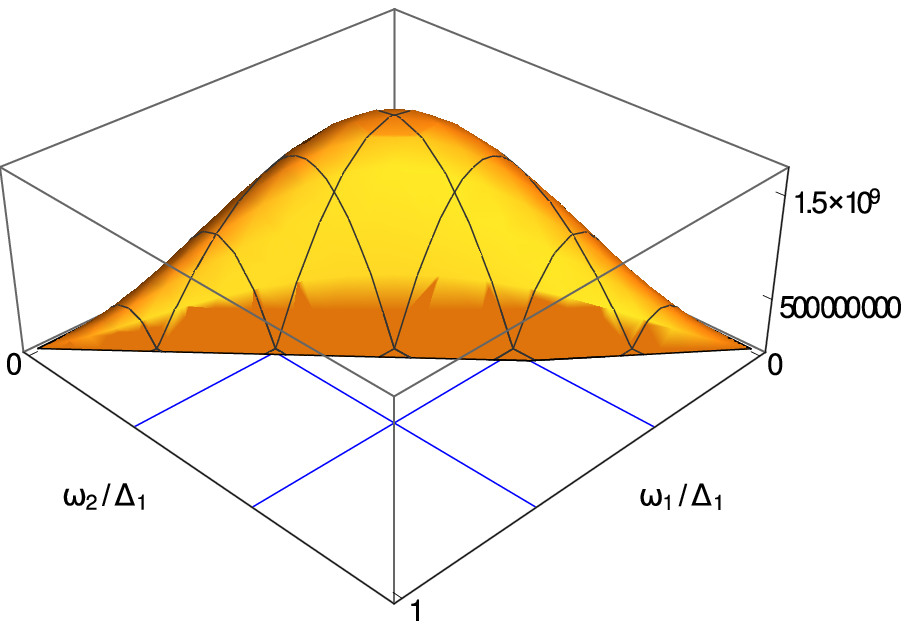}
\end{figure}

\begin{table}
\caption{Allowed values of the total electron momentum $ J_e $ for the systems of different numbers $ N_e $ of equivalent electrons with angular momenta $ j_e $
within $ jj $-coupling scheme. The results are given only for the even numbers $ N_e $, since for odd $ N_e $ the $ J_e $ values are half-integer and no analogy with photons can be traced. The numbers in parenthesis show how many times a particular value $J_e$ can arrive among the allowed values. The maximum number of equivalent electrons with particular value $ j_e $ is $ 2j_e+1 $. The allowed values of $ J_e $ are the same for configurations with $ N_e $ and $ 2j_e+1-N_e $ electrons, respectively.}
\begin{tabular}{| c | c | c |}
\hline\hline
 $ j_e $  & $ N_e $ & $ J_e $
\qquad\\
\hline $ 1/2 $ & $ 2 $ & $ 0 $
\qquad\\
\hline $ 3/2 $ & $ 2 $ & $ 0,2 $
\qquad\\
\hline $ 5/2 $ & $ 2,4 $ & $ 0,2,4 $
\qquad\\
\hline $ 7/2 $ & $ 2,6 $ & $ 0,2,4,6 $
\qquad\\ $  $  & $ 4 $   & $ 0, 2(2), 4(2), 5, 6, 8 $
\qquad\\
\hline
\end{tabular}
\end{table}

\end{document}